\documentclass{aa}  

\usepackage{graphicx}
\usepackage{xcolor}
\usepackage[colorlinks=true,citecolor=blue,linkcolor=blue,urlcolor=blue,breaklinks=true]{hyperref} 

\usepackage{txfonts}
\usepackage{gensymb}

\def\refbf#1{#1}

\begin{document}

\title{New upper limits on low-frequency radio emission \\ from isolated neutron stars with LOFAR}
\titlerunning{Upper limits on radio emission from INSs with LOFAR}

   \author{I. Pastor-Marazuela \inst{1,2}
          \and
          S. M. Straal \inst{3}
          \and
          J. van Leeuwen \inst{2}
          \and
          V. I. Kondratiev  \inst{2}
   }

   \institute{Anton Pannekoek Institute for Astronomy, University of Amsterdam, Science Park 904, PO Box 94249, 1090 GE Amsterdam, The Netherlands\\
              \email{ines.pastormarazuela@uva.nl}
        \and
            ASTRON, the Netherlands Institute for Radio Astronomy, PO Box 2, 7790 AA Dwingeloo, The Netherlands
        \and
            NYU Abu Dhabi, PO Box 129188, Abu Dhabi, United Arab Emirates
             }

 
\abstract{Neutron stars that show X-ray and $\gamma$-ray pulsed emission must, somewhere in the magnetosphere, generate electron-positron pairs. Such pairs are also required for radio emission, but then why do a number of these sources appear radio quiet? Here, we carried out a deep radio search towards four such neutron stars that are isolated  X-ray/$\gamma$-ray pulsars but for which no  radio pulsations have  been detected yet. These sources are 1RXS~J141256.0+792204 (Calvera), PSR~J1958+2846, PSR~J1932+1916 and SGR~J1907+0919.
Searching at lower radio frequencies, where the radio beam is thought to be wider, increases the chances of detecting these sources, compared to the earlier higher-frequency searches.
We thus carried a search for periodic and single-pulse radio emission with the LOFAR radio telescope at 150 MHz.
We used the known periods, and searched a wide range of dispersion measures, as the distances are not well constrained.
We did not detect pulsed emission from any of the four sources. However, we put very constraining upper limits on the radio flux density at 150 MHz, of $\lesssim 1.4$ mJy.}

   \keywords{Stars: neutron -- pulsars: general}

   \maketitle
%

\section{Introduction}

Through their spin and magnetic field,
neutron stars act as powerful cosmic dynamos that can generate a wide variety of electromagnetic emission.
There thus exist many subclasses of neutron stars, with different observed behavior.
The evolutionary links between some of the classes are established, while for others these connections are currently unknown. 
The largest group in this varied population is formed by the regular rotation-powered radio pulsars.
The fast spinning, high magnetic field influx to this group are the young pulsars.
These show a high spin-down energy loss rate $\dot{E}$, and a number of energetic phenomena such as radio giant pulse (GP) emission.
The most extreme of these fast-spinning and/or high-field sources could potentially also power Fast Radio Bursts \citep[FRBs; e.g.][]{pastor-marazuela_fast_2022}.
On the long-period outskirts of the $P$-$\dot{P}$ diagram, slowly-rotating pulsars \citep[e.g.][]{young_radio_1999, 2018ApJ...866...54T} and magnetars \citep[e.g.][]{caleb_discovery_2022, hurley-walker_radio_2022} sometimes continue to shine.


\refbf{Some neutron stars, however, only shine intermittently at radio frequencies. The rotating radio transients (RRATs) burst very irregularly, and in the $P$-$\dot{P}$ diagram most are found near the death line \citep{2011MNRAS.415.3065K}, between the canonical radio pulsars and magnetars. The exact evolutionary connection between RRATs and the steadily radiating normal pulsars is unclear, but studies suggest the presence of an evolutionary link between these different classes \citep[e.g.][]{burke-spolaor_rotating_2012}.}

Finally, populations of neutron stars exist that appear to not emit in radio at all: radio-quiet magnetars such as most anomalous X-ray pulsars (AXPs) and soft gamma repeaters (SGRs), X-ray dim isolated neutron stars (XDINSs; \citealt{2007Ap&SS.308..181H}), and $\gamma$-ray pulsars \citep[e.g.][]{2013ApJS..208...17A}. These are able to produce high-energy emission but are often radio quiet.
\refbf{\citep{gencali_long-term_2018} proposed RRATs can evolve into XDINSs through a fallback accretion disk, thus becoming radio quiet.}
\refbf{However, the magnetar SGR~1935+2154 was recently seen to emit a bright radio burst bridging the gap in radio luminosities between regular pulsars and FRBs \citep{chimefrb_collaboration_bright_2020, bochenek_fast_2020, 2022ATel15697....1M}. This suggests magnetars could explain the origin of some, if not all, extragalactic FRBs.}

Potentially, some of these could produce radio emission only visible at low radio frequencies.
Detections of radio pulsations of the $\gamma$ and X-ray pulsar \emph{Geminga}, PSR~J0633+1746, have been claimed at and below the 100\,MHz observing frequency range \citep[][]{malofeev_detection_1997,malov_geminga_2015,2015ApJ...815..126M}, although a very deep search using the low frequency array \citep[LOFAR][]{van_haarlem_lofar_2013} came up empty \cite[Ch.~6 in][]{coen13}.
Such low-frequency detections offer an intriguing possibility to better understand the radio emission mechanism of these enigmatic objects.
Radio detections of a magnetar with LOFAR, complementary to higher-frequency studies such as \citet{camilo_transient_2006} and \citet{2022ApJ...931...67M} for  XTE~J1810$-$197, could offer insight into emission mechanisms and propagation in ultra-strong magnetic fields.

XDINSs feature periods that are as long as those in magnetars, but they display less extreme magnetic field strength. 
The XDINSs form a small group of seven isolated neutron stars that  show thermal emission in the soft X-ray band.
Since their discovery with ROSAT in the 1990s, several attempts were made to detect these sources at radio frequencies, but they were unsuccessful \citep[e.g.][]{kondratiev_new_2010}. As those campaigns operated above 800\,MHz, a sensitive lower-frequency search could be opportune. It has been proposed \citep[e.g.][]{komesaroff_possible_1970,cordes_observational_1978} and observed \citep[e.g.][]{chen_frequency_2014} that pulsar profiles are usually narrower at higher frequencies and become broader at lower radio frequencies.
This suggests the radio emission cone is broader at low frequencies, and sweeps across a larger fraction of the sky as seen from the pulsar.
\refbf{Additionally, radio pulsars often present negative spectral indices, and are thus brighter at lower frequencies \citep{bilous_lofar_2016}}.
\refbf{If all neutron star radio beams are broader and brighter at lower frequencies,} chances of detecting radio emission from $\gamma$ and X-ray Isolated Neutron Stars (INSs) increase at the lower radio frequencies offered through LOFAR. The earlier observations that resulted in non-detections could then have just missed the narrower high-frequency beam, where the wider lower-frequency beam may, in contrast, actually enclose Earth. In that situation, LOFAR could potentially detect the source.

Recently, a number of radio pulsars were discovered that shared properties with XDINSs and RRATs, such as soft X-ray thermal emission, a similar position in the $P$-$\dot{P}$ diagram, and a short distance to the solar system. These sources, PSR~J0726$-$2612 \citep{2019A&A...627A..69R} and PSR~J2251$-$3711 \citep{2020MNRAS.493.1165M}, support the hypothesis that XDINS are indeed not intrinsically radio quiet, but have a radio beam pointed away from us.
These shared properties could reflect a potential link between the radio and X-ray emitting pulsars with XDINSs and RRATs.
A firm low-frequency radio detection of INSs would thus tie together these observationally distinct populations of neutron stars.

In this work we present LOFAR observations of \refbf{four} INSs that brightly pulsate at X-ray or $\gamma$-ray energies, but have not been detected in radio. These sources are listed in Section~\ref{sec:sources}, and their parameters are presented in Table~\ref{tab:setup}.


\section{Targeted sources} \label{sec:sources}

\subsection{J1412+7922}
The INS 1RXS~J141256.0+792204, dubbed "Calvera" and hereafter J1412+7922, was first detected with \textit{ROSAT} \citep{voges_rosat_1999} as an X-ray point source, and subsequently with \textit{Swift} and \textit{Chandra} \citep{rutledge_discovery_2008, shevchuk_chandra_2009}. X-ray observations confirmed its neutron star nature through the detection of $P\simeq59$\,ms pulsations by \cite{zane_discovery_2011}, and allowed for the determination of its spin-down luminosity $\dot{E}\sim6\times10^{35}$ erg s$^{-1}$, characteristic age $\tau_c\equiv P/2\dot{P}\sim3\times10^5$ years, and surface dipole magnetic field strength $B_s=4.4\times10^{11}$ G by \cite{halpern_x-ray_2013}. Although these values are not unusual for a rotationally-powered pulsar, the source is not detected in radio \citep{hessels_strong_2007,zane_discovery_2011} or  $\gamma$-rays \citep{2021ApJ...922..253M}.
The X-ray emission can be modelled with a two-temperature black body spectrum \citep{zane_discovery_2011}, similar to other XDINS \citep{pires_xmm-newton_2014}.
However, J1412+7922 shows a spin period much faster than typically observed in XDINS. Since the source is located at high galactic latitudes and its inferred distance  is relatively low \citep[$\sim$3.3\,kpc;][]{2021ApJ...922..253M} the path through the interstellar medium is not long enough to explain the radio non-detections by high \refbf{dispersion measure} (DM) or scattering values.

\subsection{J1958+2846} 
\label{sec:intro:J1958}
Discovered by \cite{abdo_detection_2009} through a blind frequency search of \textit{Fermi-LAT} $\gamma$-ray data, INS PSR~J1958+2846, hereafter J1958+2846, has shown no X-ray or radio continuum emission counterpart so far \citep{ray_precise_2011,frail_known_2016}.
Arecibo observations have put very constraining upper limits of 0.005\,mJy at 1510\,MHz \citep{ray_precise_2011}.
Searches for pulsations from the source using the single international LOFAR station FR606 by \citet{2021A&A...654A..43G} also found no periodic signal. 

The double-peaked pulse profile of J1958+2846 can be interpreted as a broad $\gamma$-ray beam. The earlier higher-frequency radio non-detections could be due to a narrower radio beam and to an unfavourable rotation geometry with respect to the line of sight. If the radio beam is indeed wider at lower frequencies, LOFAR would have higher chances of detecting it. In that case, a setup more sensitive than the \citet{2021A&A...654A..43G} single-station search is required.

Modeling by \cite{pierbattista_light-curve_2015} indicates that the $\gamma$-ray pulse profile of J1958+2846 can be well fitted by One Pole Caustic emission (OPC, \citealt{romani_constraining_2010}, \citealt{watters_atlas_2009}) or an Outer Gap model (OG, \citealt{cheng_three-dimensional_2000}).
In both cases,  the $\gamma$-rays are generated at high altitudes above the NS surface. Each model constrains the geometry of the pulsar. For the OPC model, the angle between the  rotation and  magnetic axes $\alpha = 49\degr$, while the angle between the observer line-of-sight and the rotational axis $\zeta = 85\degr$. 
The OG model reports similarly large angles, with the NS equator rotating in the plane that also contains Earth, and an  oblique dipole: $\alpha=64\degr,~  \zeta = 90\degr$. 
If this model is correct, the low-frequency radio beam would thus need to be wider than $\sim$30\degr\ to encompass the telescope. 
\refbf{That is uncommonly wide; only 8 out of the 600 pulsars in the ATNF catalogue that are not recycled and have a published 400\,MHz flux, 
have a duty cycle suggestive of a beam wider than 30\% \citep{manchester_atnf_2005}.}
As such a width is unlikely, a total-intensity detection would thus suggest to first order a geometry where $\alpha$ and $\zeta$ are closer than follows from  \cite{pierbattista_light-curve_2015}, even if  that suggestion would only be qualitative.
Subsequent follow-up measurements of polarisation properties throughout the pulse, and fitting these to the rotating vector model \citep[RVM;][]{radhakrishnan_magnetic_1969}, can quantify allowed geometries to within a relatively precise combinations of $\alpha$ and $\zeta$. 
As a matter of fact, in a similar study on radio-loud $\gamma$-ray pulsars, \citet{2015MNRAS.446.3367R} already find that RVM fits suggest that the magnetic inclination angles $\alpha$ are much lower than predicted by the $\gamma$-ray light curve models.
This, in turn, affirms that deep radio searches can lead to detections even when the $\gamma$-ray light curves suggest the geometry is unfavorable.


\subsection{J1932+1916} 
The INS PSR~J1932+1916, hereafter J1932+1916, was discovered in \textit{Fermi-LAT} data through blind searches with the $Einstein@Home$ volunteer computing system \citep{clark_einsteinhome_2017}. J1932+1916 is the youngest and  $\gamma$-ray brightest among the four $\gamma$-ray pulsars presented from that effort in \citep{pletsch_einsteinhome_2013}.
The period is 0.21\,s, the characteristic age is 35\,kyr.
\refbf{\citet{frail_known_2016} find no continuum 150\,MHz source at this position with GMRT at a flux density upper limit of 27\,mJy\,beam$^{-1}$, with $1\sigma$ errors. If the flux density they find at the position of the pulsar is in  fact the pulsed emission from J1932+1916, then a LOFAR periodicity search as described here should detect the source at a S/N of 15 if the duty cycle is 10\%.}
\cite{karpova_observations_2017} report on a potential pulsar wind nebula (PWN) association from \textit{Swift} and \textit{Suzaku} observations. However, no X-ray periodicity searches have been carried out before.

\subsection{J1907+0919}
The Soft Gamma Repeater J1907+0919, also known as SGR 1900+14, was detected through its bursting nature by \cite{mazets_soft_1979}. Later outbursts were detected in 1992 \citep{kouveliotou_recurrent_1993}, 1998 \citep{hurley_asca_1999} and 2006 \citep{mereghetti_first_2006}. The August 1998 outburst allowed the detection of an X-ray period of $\sim5.16$ s, and thus confirmed the nature of the source as a magnetar \citep{hurley_asca_1999,kouveliotou_discovery_1999}.
\cite{frail_relativistic_1999} detected a transient radio counterpart that appeared simultaneous to the 1998 outburst, and they identified the radio source as a synchrotron emitting nebula.
\cite{shitov_low_2000} claimed to have found radio pulsations at 111\,MHz from four to nine months after the 1998 burst, \refbf{but the number of trials involved in the search, the small bandwidth of the system, and the low S/N of the presented plots, lead us to conclude the confidence level for these detections is low.} 
No other periodic emission has been found at higher radio frequencies \citep{lorimer_psr_2000,fox_high-resolution_2001,lazarus_constraining_2012}.\\

This paper is organised as follows: in Section~\ref{sec:observations} we explain how we used  LOFAR \citep{van_haarlem_lofar_2013} to observe the sources mentioned above; in Section~\ref{sec:data} we detail the data reduction procedure, including the periodicity and the single pulse searches that we carried; in Section~\ref{sec:results} we present our results, including the upper limit that we set on the pulsed emission; in Section~\ref{sec:discussion} we discuss the consequences of these non-detections for the radio-quiet pulsar population, and in Section~\ref{sec:conclusion} we give our conclusions on this work.

\section{Observations} \label{sec:observations}
We observed the four sources with the largest possible set of High Band Antennas (HBAs) that LOFAR can coherently beam form. 
Each observation thus added 22 HBA Core Stations, covering 78.125\,MHz bandwidth in the 110 MHz to 190 MHz frequency range (centered on 148.92\,MHz),
with 400 channels of 195\,kHz wide.
The LOFAR beam-forming abilities allow us to simultaneously observe different regions of the sky \citep{ls10,stappers_observing_2011, coenen_lofar_2014}.
For our point-source searches of INSs, we used three beams per observation; one beam pointed to the source of interest, one on a nearby known pulsar, and one as a calibrator blank-sky beam to cross-check potential candidates as possibly arising from Radio Frequency Interference (RFI).
We carried out observations between 16 January 2015 and 15 February 2015 under project ID 
LC3\_036\footnote{After we completed the current manuscript as \citet[][PhD Thesis, Ch.~2]{pastor-marazuela_exploring_2022}, \citet{arias_possible_2022} posted a pre-print presenting partly the same data.}.
We integrated for 3 hours on each of our sources. The data was taken in Stokes I mode.
Since the periods of the $\gamma$-ray pulsars are known, the time resolution of each observation was chosen such to provide good coverage of the pulse period, at a sampling time between 0.16$-$1.3\,ms. The observation setup is detailed in Table~\ref{tab:setup}.

\begin{table*}
  \caption{Parameters of the observed pulsars and observational setup of the observations in the LC3\_036 proposal. The beam of each observation was centered in the reported pulsar coordinates.
  Listed in the bottom rows are the earlier periodicity and single pulse search limits. The upper limits from \citet{frail_known_2016} described in the main text are period-averaged flux densities and are not listed here. The last row lists  the limits from the current work, for S/N=5, with errors of 50\% \citep{bilous_lofar_2016}.  }             
\label{tab:setup}     
\centering                          
\begin{tabular}{lp{2.6cm}p{2.6cm}p{2.6cm}p{2.6cm}}
\hline\hline \\ [-1.5ex]
& J1412+7922 & J1958+2846 & J1932+1916 & J1907+0919 \\
\hline \\ [-1ex]
Right ascension, $\alpha$ (J2000)\dotfill & 14 12 56 & 19 58 40  & 19 32 20 & 19 07 14.33 \\
Declination, $\delta$ (J2000)    \dotfill & +79 22 04 & +28 45 54 & +19 16 39 & +09 19 20.1 \\
Period, $P$ (s)                  \dotfill & 0.05919907107 & 0.29038924475 & 0.208214903876 & 5.198346 \\
Period derivative, $\dot{P}$ (s s$^{-1}$)\dotfill & 3.29134$\times10^{-15}$ &  2.12038$\times10^{-13}$ & 9.31735$\times10^{-14}$ & 9.2$\times10^{-11}$ \\
Epoch (MJD)            \dotfill & 58150$^a$  & 54800$^b$ & 55214$^c$ & 53628$^d$  \vspace{1ex}\\
LOFAR ObsID            \dotfill &L257877 & L258545 & L259173 & L216886 \\
 Obs. date (MJD)        \dotfill & 57038 & 57046 & 57068 & 56755 \\
Sample time (ms)       \dotfill & 0.16384 & 1.31072 & 1.31072 & 0.65536 \\
Test pulsar detected   \dotfill & B1322+83 &  B1952+29 & B1933+16 & B1907+10  \vspace{1ex}\\ 
Periodic flux density (mJy @ GHz)\dotfill& <4 @ 0.385$^e$   & <2.0 @ 0.15$^{g}$ & <2.9 @ 0.15$^{g}$  &  50 @ 0.111$^h$ \\
  & <0.05 @ 1.36$^f$  & <0.005 @ 1.51$^b$ & <0.075 @ 1.4$^c$ & <0.4 @ 0.43$^i$\\
  & <0.3  @ 1.38$^e$ & & & <0.3 @ 1.41$^i$ \\
  &&&&  <0.012 @ 1.95$^j$  \vspace{1ex}\\
LOFAR periodic sensitivity $S_{\text{lim,p}}$ (mJy)\dotfill
  & $0.26\pm0.13$ & $0.53\pm0.26$ & $0.73\pm0.36$ &   $1.39\pm0.69$
\vspace{0.5ex} \\
LOFAR single pulse sensitivity $S_{\text{lim,sp}}$ (Jy)\dotfill
  & $1.47\pm0.73$ & $1.35\pm0.68$ & $2.20\pm1.10$ &   $0.84\pm0.82$
\vspace{0.5ex} \\
\hline
\end{tabular}
\tablefoot{\footnotesize $^a$\cite{bogdanov_neutron_2019}, $^b$\cite{ray_precise_2011}, $^c$\cite{pletsch_einsteinhome_2013}, $^d$\cite{mereghetti_first_2006}, $^e$\cite{hessels_strong_2007}, $^f$\cite{zane_discovery_2011},  $^g$\cite{2021A&A...654A..43G}, $^h$\cite{shitov_low_2000},
$^i$\cite{lorimer_psr_2000}, $^j$\cite{lazarus_constraining_2012}}
\end{table*}

\section{Data reduction} \label{sec:data}

The data was pre-processed by the LOFAR pulsar pipeline after each observation \citep{alexov_lofar_2010, stappers_observing_2011} and stored on the LOFAR Long Term Archive\footnote{LTA: \url{https://lta.lofar.eu/}} in \texttt{PSRFITS} format \citep{hotan_psrchive_2004}. 
The 1.5 TB of data was then transferred to one of the nodes of the Apertif real-time FRB search cluster \emph{ARTS}
\citep{van_leeuwen_arts_2014, van_leeuwen_apertif_2022}. 

We performed a periodicity search as well as a single-pulse search using \textsc{Presto}\footnote{\textsc{Presto}: \url{https://www.cv.nrao.edu/~sransom/presto/}} \citep{ransom_new_2001}.
The data was \refbf{cleaned of RFI using first \texttt{rfifind}, and then removing impulsive and periodic signals at DM=0~pc~cm$^{-3}$. Next we searched the clean data for periodic signals and single pulses.}
We searched for counterparts around the known $P$ and $\dot{P}$ of each pulsar. Additionally, we performed a full blind search in order to look for potential pulsars in the same field of view, since many new pulsars are found at low frequencies \citep{sanidas_lofar_2019} and chance discoveries happen regularly (e.g., \citealt{2020MNRAS.492.4825O}).
Since the DM of our sources is unknown, we searched over a range of DMs going from 4 pc cm$^{-3}$ to 400 pc cm$^{-3}$. The DM-distance relation is not precise enough to warrant a much smaller DM range, even for sources for which a distance estimate exists; and a wider DM range allows for discovery of other pulsars contained in our field of view.
The highest DM pulsar detected with LOFAR has a DM = 217 pc cm$^{-3}$ \citep{sanidas_lofar_2019}. We thus searched up to roughly twice this value to make sure that any detectable sources were covered. We determined the optimal de-dispersion parameters with \texttt{DDplan} from \textsc{Presto}.
The sampling time variation between some of the four observations had a slight impact on the exact transitions of the step size but generally the data was de-dispersed in steps of 0.01 pc cm$^{-3}$ up to DM~=~100~pc~cm$^{-3}$; then by 0.03~pc~cm$^{-3}$ steps up to 300~pc~cm$^{-3}$ and finally using 0.05~pc~cm$^{-3}$ steps.

We manually inspected all candidates down to $\sigma=4$, resulting in $\sim$$1400$ candidates per beam. 
To verify our observational setup, we performed the same blind search technique to our test pulsars B1322+83 and B1933+16, which we detected. The test pulsar B1953+29 was not detected because the sampling time of the observation of J1958+2846 was not adapted to its $\sim6$ ms period.
However, we were able to detect B1952+29 \citep{hewish_observation_1968} in this same pointing. Even though it is located at >1\degree{} from the targeted coordinates, it is bright enough to be visible as a side-lobe detection.

The candidates from \textsc{Presto}'s single pulse search were further classified using the deep learning classification algorithm developed by \cite{connor_applying_2018}, which has been verified and successful in the Apertif surveys \citep[e.g.][]{connor_bright_2020,pastor-marazuela_chromatic_2021}.  This reduced the number of candidates significantly by sifting out the remaining RFI. The remaining candidates were visually inspected.

\section{Results} \label{sec:results}
In our targeted observations we were unable to detect any plausible astronomical radio pulsations or single pulses. We determine new 150 MHz flux upper limits by computing the sensitivity limits of our observations.
%
To establish these sensitivity limits, we apply the radiometer equation adapted to pulsars, detailed below.
We determine the telescope parameters that are input to this equation by following the procedure\footnote{\url{https://github.com/vkond/LOFAR-BF-pulsar-scripts/blob/master/fluxcal/lofar_fluxcal.py}} described in \cite{kondratiev_lofar_2016} and \cite{mikhailov_lofar_2016}.
That approach takes into account the system temperature (including the sky temperature), the projection effects governing the effective area of the fixed tiles, and the amount of time and bandwidth removed due to RFI, to produce the overall observation system-equivalent flux density (SEFD).\\

\refbf{
For the sensitivity limit on the periodic emission we use 
the following equation (see., e.g., \citealt{1985ApJ...294L..25D}): 
\begin{equation} \label{eq:Sp}
S_{\text{lim,p}} = \beta \dfrac{T_{\text{sys}}}{G \sqrt{n_{\text{p}}\ \Delta\nu\ t_{\text{obs}}}} \times S/N_{\text{min}} \times \sqrt{\dfrac{W}{P-W}},    
\end{equation}
where $\beta\lesssim1$ is a digitisation factor, $T_{\text{sys}}$\,(K) is the system temperature, $G$\,(K\,Jy$^{-1}$) is the telescope gain, $\Delta\nu$\,(Hz) is the observing bandwidth, and $t_{\text{obs}}$\,(s) is the observation time. $P$\,(s) represents the spin period, while $W$\,(s) gives the pulsed width assuming a pulsar duty cycle of 10\%.}
To facilitate direct comparison \refbf{of the periodic emission limits} to values reported in e.g., \citet{ray_precise_2011} and \citet{2021A&A...654A..43G}, we use a minimum signal-to-noise ration $S/N_{\text{min}}=5$. 
A more conservative option, given the high number of candidates per beam, would arguably be to use a limit of S/N=8. We did, however, review by eye all candidates with S/N>4; and the reader can easily scale the reported sensitivity limits to a different S/N value.

\refbf{
The sensitivity limit on the single pulse emission, $S_{\text{lim,sp}}$, is computed as follows:
\begin{equation} \label{eq:Ssp}
    S_{\text{lim,sp}} = \beta \dfrac{T_{\text{sys}}}{G \sqrt{n_{\text{p}}\ \Delta\nu\ t_{\text{obs}}}} \times S/N_{\text{min}} \times \sqrt{\dfrac{t_{\text{obs}}}{W}},
\end{equation}
where all variables are the same as in Equation~\ref{eq:Ssp}. We searched for single pulses down to a signal-to-noise ratio $S/N_{\text{min}}=7$.
}

We report these \refbf{periodic and single pulse} sensitivity limits, computed at the coordinates of the central beam of each observation, in Table~\ref{tab:setup}.
Even though all observations are equally long,  the estimated $S_{\text{lim,p}}$ values are different.
That is mostly due to the strong dependence of the LOFAR effective area, and hence
the sensitivity, on the elevation.

\begin{figure}
    \centering
    \includegraphics[width=\hsize]{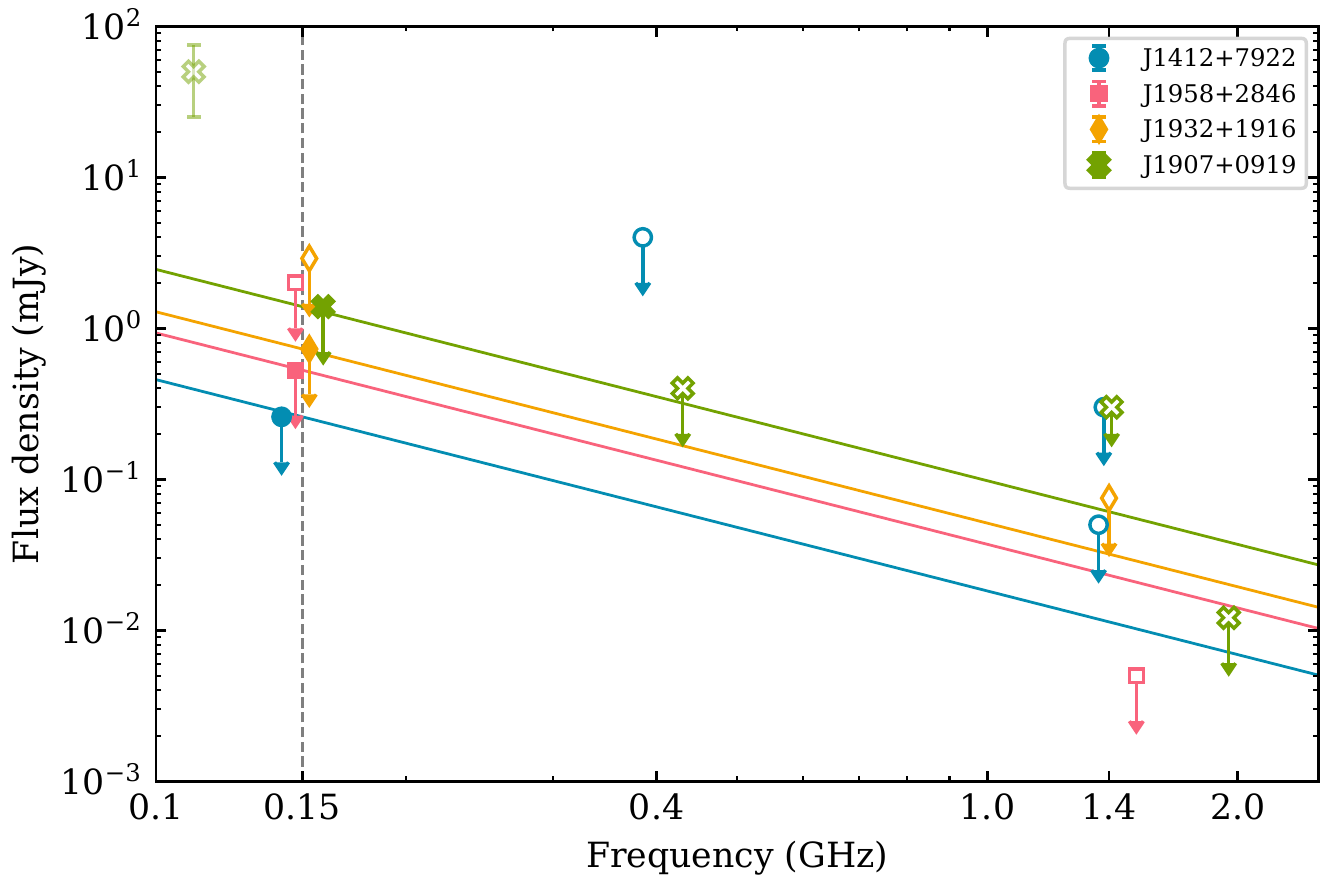}
    \caption{Flux density upper limits of this work at 150 MHz
      (filled symbols) with S/N~=~5 for comparison to earlier
      searches of the same sources 
      (empty symbols). Solid lines going through our upper limit estimates with spectral index $\alpha = -1.4$ are overlaid to show the scaling of our sensitivity limits. Our limits are plotted slightly offset from the 150 MHz observing frequency (dashed line) for better visibility. \refbf{The faded green marker for SGR~J1907+0919 represents  the claimed detection from  \citet{shitov_low_2000}.}}
    \label{fig:sensitivity}
\end{figure}

In Fig.~\ref{fig:sensitivity}, we compare  our upper limits to those established in previous  searches, mostly using the same techniques. 
\refbf{Our upper limit on the flux of J1907+0919 is $\sim$50$\times$  deeper than the claimed 1998-1999 detections, at the same 3-m wavelength, with BSA \citep{shitov_low_2000}.}
Other searches were  generally undertaken at higher frequencies \citep{hessels_strong_2007, zane_discovery_2011, ray_precise_2011, pletsch_einsteinhome_2013, 2021A&A...654A..43G}.
If we assume that these four pulsars have radio spectra described by a single power-law $S_{\nu} \propto \nu^{\alpha}$ with a spectral index of $\alpha=-1.4$ \citep{bates_pulsar_2013,bilous_lofar_2016}, the  upper limits we present here for  J1412+7922 and J1932+1916 are the most stringent so-far for any search.
The upper limits on J1958+2846 (Arecibo; \citealt{ray_precise_2011}) and J1907+0919 (GBT; \citealt{lazarus_constraining_2012}) are a factor of 2--3 more sensitive than ours. 
\refbf{However, pulsars present a broad range of spectral indices. If we take the mean $\pm2\sigma$ measured by \cite{jankowski_spectral_2018}, spectral indices can vary from $-$2.7 to $-$0.5. The flux upper limits we measure would be the deepest assuming a $-$2.7 spectral index, but the shallowest at $-$0.5.}


\section{Discussion} \label{sec:discussion}

\subsection{Comparison to previous limits}

For J1958+2846 and J1932+1916, we can make a straightforward relative comparisons between  our results presented here and the existing limit at 150\,MHz, from the single-station LOFAR campaign by
\citet{2021A&A...654A..43G}. Our 22 Core Stations are each 1/4th of the area of the FR606 station and are coherently combined,
leading to a factor $\frac{A_\mathrm{core}}{A_\mathrm{FR606}}=\frac{22}{4}$ difference in area $A$ for the radiometer equation and $S_{\text{lim}}$.
The integration time $t$ of 3\,h is shorter than the FR606 total of 8.3\,h (J1958+2846) and 4.1\,h (J1932+1916), leading to
a factor $\sqrt{\frac{t_\mathrm{core}}{t_\mathrm{FR606}}}=\sqrt{\frac{3}{8.3}}$
in the radiometer equation.
Other factors such as the sky background and the influence of zenith angle on the sensitivity should be mostly the same for both campaigns. 
Our $S_{\text{lim}}$ is thus $\frac{22}{4}\sqrt{\frac{3}{8.3}}=3.3$ times deeper than the \citet{2021A&A...654A..43G} upper limit for J1958+2846, and 4.7 times for J1932+1916. 
Those factors are in good agreement with the actual limits listed in Table~\ref{tab:setup}.

\refbf{
In \cite{bilous_lofar_2016}, they measured the mean flux density $S_{\text{mean}}$ of 158 pulsars detected with LOFAR, where $S_{\text{lim,p}} = S_{\text{mean}} \times \sqrt{W/(P-W)} = S_{\text{mean}}/3$. 
Compared to those LOFAR detections, our upper limit on J1412+7922 is deeper than all 158 sources (100\%), J1958+2846 is deeper than 156 sources (99\%), J1932+1916 is deeper than 144 sources (93\%), and J1907+0919 is deeper than 109 sources (69\%). 
The flux upper limits we have set on each of the sources in our sample are some of the deepest compared to other LOFAR radio pulsar detections. Longer observing times are thus unlikely to result in a detection or improve our flux upper limits. Additional follow up would only be constraining with more sensitive radio telescopes.
}

\subsection{Emission angles and intensity}

Different pulsar emission mechanism models exist that predict radio and $\gamma$-ray emission to be
simultaneously formed in the pulsar magnetosphere. \refbf{The emission sites are not necessarily co-located, though.
The periodic radio emission is generally thought to be formed just above the polar cap.}
The high-energy polar cap (PC) model next assumes that the $\gamma$-ray emission is also produced 
near the surface of the NS, and near the magnetic polar caps.
In the outer magnetosphere emission models, such as the Outer Gap (OG) or the One Pole Caustic (OPC) models,
on the other hand, 
the $\gamma$-ray emission is produced high up in the magnetosphere of the NS, within the extent of the light cylinder.

\refbf{
For the sources in our sample, specific high-energy geometry models have only been proposed for J1958+2846 \citep{pierbattista_light-curve_2015}.
A detection could have confirmed one of these (Sect.~\ref{sec:intro:J1958}). But also for our sample in general, conclusions can be drawn from the non detections. The two general high-energy model classes mentioned above predict different, testable beam widths. Our radio non-detections, when attributed to radio beams that are not wide enough to encompass Earth, favor outer magnetospheric models \citep[see, e.g.,][]{romani_constraining_2010}. 
That is because in the OG/OPC models, the  $\gamma$-ray beam (which is detected for our sources) is much broader than the radio beam.
The radio beam, being much narrower, is unlikely cut through our line of sight. Such a model class is thus more applicable than one where the radio and high-energy beam are of similar angular size, such as the PC model (or, to a lesser extent, the slot gap model; \citealt{2003ApJ...588..430M, pierbattista_light-curve_2015}). 
In that case, detections in both radio and high-energy would be more often expected. Our results thus favor OG and OPC models over PC models for high-energy emission.
}

Note that while it is instructive to discuss the coverage of the radio pulsar beam in binary terms
-- it either  hits or misses Earth  --
this visibility is not that \refbf{unambiguous} in practice.
The beam edge is not sharp.
\refbf{In a beam mapping experiment} enabled by the geometric precession in PSR~J1906+0745 \citep{2015ApJ...798..118V},
the flux at the edge of the beam is over 100$\times$ dimmer than the peak, but it is still
present and detectable \citep{2019Sci...365.1013D}. Deeper searches thus continue to have value,
even if non-detections at the same frequency already exist.

That said, the detection of PSR~J1732$-$3131 only at 327 and potentially even 34\,MHz \citep{maan_deep_2014} shows that \refbf{emission beam widening} (or, possibly equivalently, a steep spectral index) at low frequencies is a real effect, also for $\gamma$-ray pulsars.

\subsection{Emission  mechanism and evolution}
Most models explain the radio quietness of an NS through
a chance beam misalignment, as above.
It could, of course, also be a more  intrinsic property.
There are at least two regions in the $P$-$\dot{P}$ diagram
where radio emission may be increasingly hard to generate.

The first parameter space of interest is for sources close to the radio death line \citep{1993ApJ...402..264C}.
XDINSs are preferably found there,
which suggests these sources are approaching, in their evolution,
a state in which radio emission generally ceases.
From what we see in normal pulsars,
the death line represents the transition into a state in which  electron-positron pair
formation over the polar cap completely ceases.
Once the pulsar rotates too slowly to generate a large enough potential drop over the polar cap,
required for this formation,
the radio emission turns off \citep{rs75}.
The high-energy emission also requires pair formation, but these could occur farther out.
We note that polar cap pair formation can continue at longer periods,
if the NS surface magnetic field is not a pure dipole.
With such a  decreased curvature radius, the NS may keep on shining.
Evidence for such higher-order fields is present in a number of pulsars, e.g.,
 PSR~J0815+0939 \citep{2017ApJ...845...95S} and PSR~B1839$-$04 \citep{szary_single-pulse_2020}.
This would also influence the interpretation of any polarization information,
as the RVM generally assumes a dipole field.

\begin{figure}[t]
    \centering
    \includegraphics[width=\hsize]{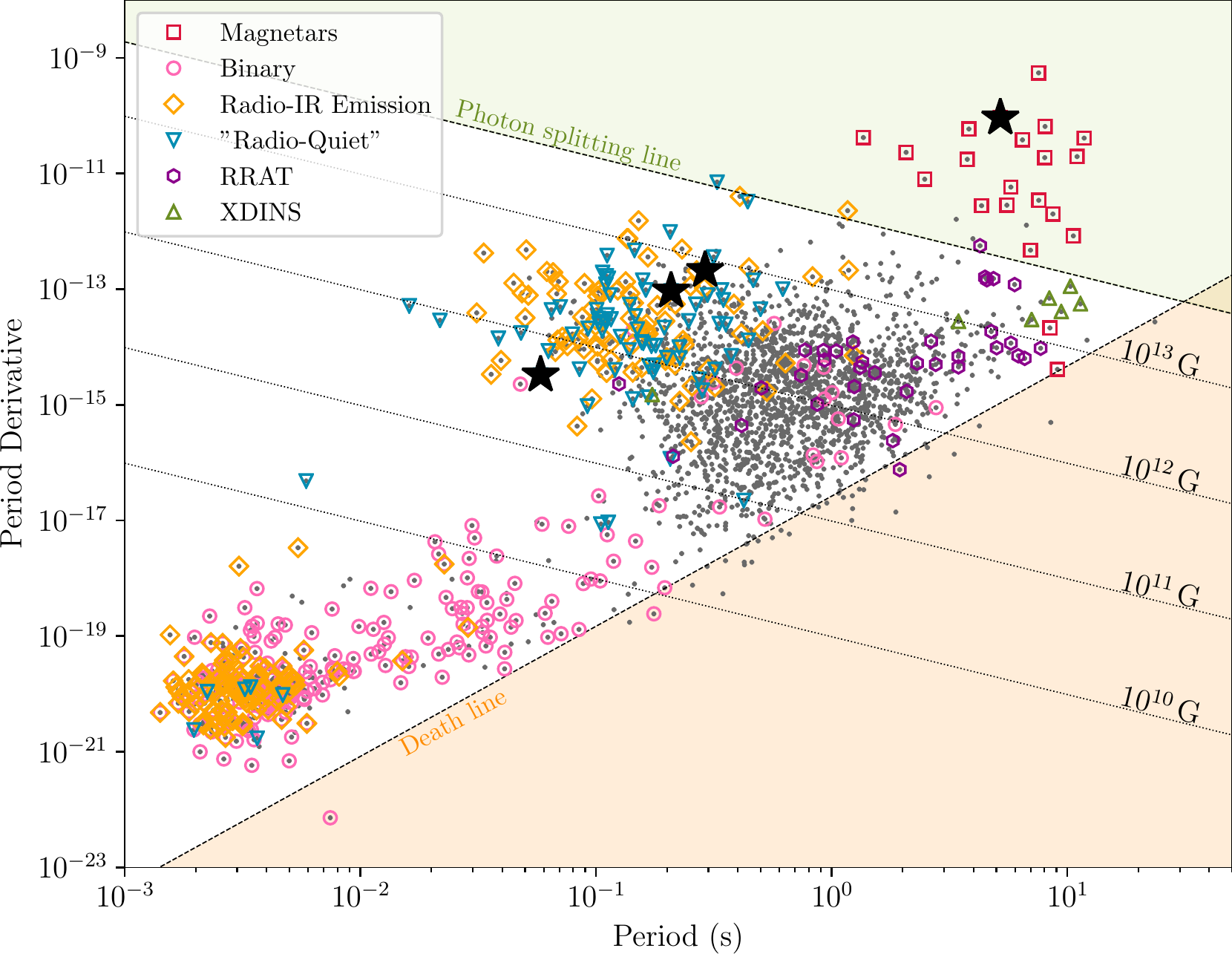}
    \caption{$P-\dot{P}$ diagram showing the location of the sources presented in this work. All pulsars from the ATNF Pulsar Catalogue \citep{manchester_atnf_2005} are shown as grey dots, with different pulsar classifications encircled by different symbols. The sources discussed in this work are shown as black stars, from left to right: J1412+7922, J1932+1916, J1932+1916, and J1907+0919. The orange shaded region is delimited by the death line, while the green shaded region is delimited by the photon splitting line. Plot generated with \texttt{psrqpy} \citep{pitkin_psrqpy:_2018}.}
    \label{fig:ppdot}
\end{figure}

None of the sources in our sample are close to this death line (See Fig.~\ref{fig:ppdot}),
but SGR~J1907+0919 is beyond a different, purported boundary:
the photon splitting line \citep{2001ApJ...547..929B}.
In pulsars in that second parameter space of interest, where magnetic fields are  stronger than the quantum critical field, of 4.4 $\times$ 10$^{13}$\,G (Fig.~\ref{fig:ppdot}), 
pair formation cannot compete with magnetic photon splitting. 
Such high-field sources could then be radio quiet but X-ray or $\gamma$-ray bright.
\refbf{We mark the critical field line for a dipole in Fig.~\ref{fig:ppdot}, 
 but note, as \citet{2001ApJ...547..929B} do, that higher multipoles and general relativistic
effects can subtly change the quiescence limit on a per-source basis.}
That said, given its  spindown dipole magnetic field strength of 7\,$\times$\,10$^{14}$\,G, 
our non-detection of SGR~J1907+0919 supports the existence of this limit.

\subsection{Propagation effects}
\refbf{
While the emission beam widening and the negative spectral index provide potential advantages when searching for pulsars at low frequencies, 
 some propagation effects such as dispersion and scattering intensify there, impeding detection of certain sources.
The largest pulsar DM detected with LOFAR is 217~pc~cm$^{-3}$, while many galactic pulsars are known to have DM>1000~pc~cm$^{-3}$. 
Although the sources studied in this work do not have radio detections and thus no known DM, 
we can estimate this DM if a hydrogen column density $N_{\text{H}}$ was  measured from soft X-ray detections. 
\citet{he_correlation_2013} find a   correlation between  $N_{\text{H}}$ and DM as follows:
$N_{\text{H}}~(10^{20}~\text{cm}^{-2}) = 0.30^{+0.13}_{-0.09}~\text{DM}~(\text{pc\ cm}^{-3})$.
}

\refbf{
While J1958+2846 and J1932+1916 have only been detected in $\gamma$-rays, J1412+7922 and J1907+0919 have soft X-ray detections where $N_\text{H}$ has been measured. 
For J1907+0919, \cite{kouveliotou_discovery_1999} measured a large $N_{\text{H}}$ value of  $3.4-5.5\times 10^{22}$~cm$^{-2}$. 
The correlation suggests a DM of 1100$-$1800~pc~cm$^{-3}$. At such a large DM the detection limit of LOFAR is severly impacted. 
Because J1907+0919 is a very slow rotator, the intra channel dispersion delay still only becomes or order 10\% of the period, which means peridiocity searches could in principle still detect it; but the flux density per bin is of course much decreased when the pulse is smeared out over 100s of time bins. }

\refbf{
In contrast, \cite{shevchuk_chandra_2009} reported a measured $N_{\text{H}}=3.1\pm0.9 \times 10^{20}$~cm$^{-2}$ for J1412+7922. We thus estimate its DM to be in the range 5--15~pc~cm$^{-3}$. This low DM would have easily been detected with LOFAR.}


\section{Conclusion} \label{sec:conclusion}

We have conducted deep LOFAR searches of periodic and single-pulse radio emission from four isolated neutron stars.
Although we validated the observational
setup with the detection of the test pulsars, we did not detect any of the four targeted pulsars.
This can be explained with an
intrinsic radio-quietness of these sources, as was previously proposed.
It could also be caused by a chance misalignment between the radio beam and the line of sight.

With the new upper limits, we can rule out the hypothesis that INSs had not been previously detected at
radio frequencies around 1\,GHz, because of a steeper spectrum than that of regular radio pulsars.
\refbf{Since radio emission from magnetars has been detected after high energy outbursts \cite[e.g.][]{2022ATel15697....1M}, additional radio observations of J1907+0919 if the source reactivates might be successful at detecting single pulse or periodic emission in the future.}

\begin{acknowledgements}
This research was supported by 
the Netherlands Research School for Astronomy (`NOVA5-NW3-10.3.5.14'),
the European Research Council under the European Union's Seventh Framework Programme
(FP/2007-2013)/ERC Grant Agreement No. 617199 (`ALERT'), 
and by Vici research programme `ARGO' with project number
639.043.815, financed by the Dutch Research Council (NWO). 
We further acknowledge funding from National Aeronautics and Space Administration (NASA) grant number NNX17AL74G issued through the NNH16ZDA001N Astrophysics Data Analysis Program (ADAP) to SMS.
This paper is based (in part) on data obtained with the International LOFAR Telescope (ILT)
under project code
LC3\_036 (PI: van Leeuwen).
LOFAR \citep{van_haarlem_lofar_2013} is the low frequency array designed and constructed by ASTRON. It has observing, data processing, and data storage facilities in several countries, that are owned by various parties (each with their own funding sources), and that are collectively operated by the ILT foundation under a joint scientific policy. The ILT resources have benefitted from the following recent major funding sources: CNRS-INSU, Observatoire de Paris and Université d'Orléans, France; BMBF, MIWF-NRW, MPG, Germany; Science Foundation Ireland (SFI), Department of Business, Enterprise and Innovation (DBEI), Ireland; NWO, The Netherlands; The Science and Technology Facilities Council, UK; Ministry of Science and Higher Education, Poland.
\end{acknowledgements}

\bibliographystyle{yahapj} 
\bibliography{biblio} 

\end{document}